\documentclass[apj]{aastex61} %

\usepackage{graphicx,natbib}
\usepackage{amsmath}
\usepackage[T1]{fontenc}
\usepackage{txfonts}
\usepackage{changebar}
\usepackage{booktabs}
\usepackage{mathrsfs}

\shorttitle{Formation of mini-Neptunes}
\shortauthors{Venturini \& Helled}

\newcommand{\ME}{M$_{\oplus}$} 
\newcommand{\RE}{R$_{\oplus}$} 
\newcommand{\Mearth}{\text{M}_{\oplus}} 
\newcommand{\Msun}{M$_{\odot}$ }

\newcommand{\gcmcuad}{g/cm$^2$}

\newcommand{\Mcore}{M_{\text{core}}}
\newcommand{\Menv}{M_{\text{env}}}

\newcommand{\MHHe}{M_{\text{HHe}}}

\newcommand{\Mp}{M_{\text{P}}}
\newcommand{\Miso}{M_{\text{iso}}}

\newcommand{\fhhe}{f_{\text{HHe}} }

\newcommand{\tcross}{t_{\text{cross}}}
\newcommand{\tdisk}{\tau_{\text{disk}}}

\begin{document}

\title{The Formation of mini-Neptunes}
\correspondingauthor{J. Venturini}
\email{julia@physik.uzh.ch}

\author{Julia Venturini} 
\affil{Center for Theoretical Astrophysics and Cosmology, Institute for Computational Science, University of Zurich \\
Wintherthurerstrasse 190, 8057, Zurich, Switzerland}

\author{Ravit Helled}
\affil{Center for Theoretical Astrophysics and Cosmology, Institute for Computational Science, University of Zurich \\
Wintherthurerstrasse 190, 8057, Zurich, Switzerland}

\begin{abstract}
Mini-Neptunes seem to be common planets. 
In this work we investigate the possible formation histories and predicted occurrence rates of mini-Neptunes assuming the planets form beyond the iceline. We consider pebble and planetesimal accretion accounting for envelope enrichment and two different opacity conditions.     
We find that the formation of mini-Neptunes is a relatively frequent output when envelope enrichment by volatiles is included, and that there is a "sweet spot" for mini-Neptune formation with a relatively low solid accretion rate of $\sim 10^{-6}$ \ME/yr. This rate is typical for low/intermediate-mass protoplanetary disks and/or disks with low metallicities. 
With pebble accretion, envelope enrichment and high opacity favor the formation of mini-Neptunes, with more efficient formation at large semi-major axes ($\sim30$ AU) and low disk viscosity. 
For planetesimal accretion, such planets can form also without enrichment, with the opacity being a key aspect in the growth history and favorable formation location. 
Finally, we show that the formation of Neptune-like planets remains a challenge for planet formation theories.

\end{abstract}


\section{Introduction} \label{intro}
\begin{figure}
\begin{center}
	\includegraphics[width=0.7\columnwidth]{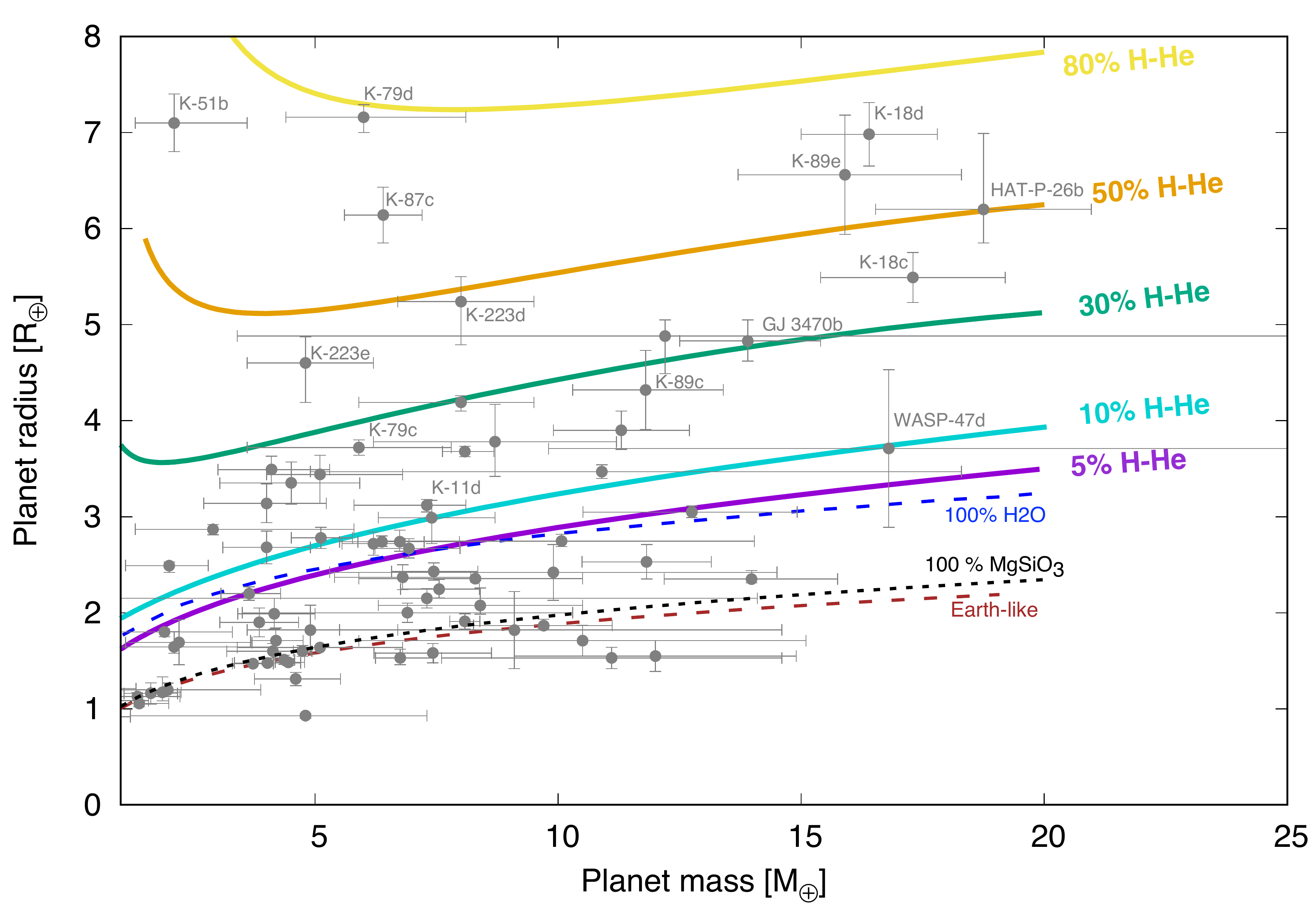}
	\caption{The mass-radius (M-R) diagram for small- and intermediate- mass exoplanets with well determined masses and radii. Also shown are M-R curves for various compositions such as water \citep[blue-dashed,][]{Lissauer11} and  ``Earth-like'' \citep[brown-dashed,][]{Dressing15}. The curves for compositions that include H-He are calculated assuming the planet has a rocky core and an envelope of H-He and water, with the amount of water being the same as that of rocks (see Sect.\ref{method}).}
\label{MR}
\end{center}
\end{figure}

The {\it Kepler} mission has revealed that planets with radii between that of the Earth (\RE) and Neptune (4 \RE) are extremely common in our galaxy \citep[e.g.,][]{Batalha13}.   
Figure \ref{MR} shows the mass-radius (M-R) relation for various compositions and many of the detected exoplanets with a relatively accurate mass determination. 
As can be seen from the figure, several of the planets have radii larger than the ones expected for pure water, suggesting that they consist of lighter elements, presumably H-He, including the low-mass planets ($ \Mp \lesssim $ 10 \ME). These planets are often referred as mini-Neptunes.

Several studies have been dedicated for the characterization of such planets confirming the existence of H-He atmospheres \citep[e.g.,][]{Weiss14, Rogers15,Dorn17a}. For example, \citet{Rogers15} showed that most of the planets with radius larger than 1.6 \RE \, are non-rocky, having volatile and/or hydrogen and helium (H-He) atmospheres. Similarly, models by \citet{Lopez12} find that the planets Kepler-11d and Kepler-11f consist of $\sim$ 30 \% of H-He. Recently, \citet{Fulton17} presented a bi-modality in the Kepler exoplanets' radii, with peaks at 1.3 and 2.4 \RE. Exoplanets with sizes clustered around 2.4 \RE \, are thought to have considerable gaseous envelopes.

Although mini-Neptunes do not exist in our Solar System, their frequent occurrence around other stars demands an explanation. In this paper we investigate the formation mechanism and occurrence rate of mini-Neptunes in the core accretion scenario. In this model, the formation of a planet begins with the growth of a solid core (composed of heavy elements) followed by the binding of a gaseous envelope. When the mass of heavy-element is comparable to the mass of H-He, known as crossover mass (which happens at time $\tcross$), a very rapid stage of gas accretion takes place, and the planet becomes a gas giant \citep{BP86, P96, HelledPPVI} if the gaseous disk is still present. When the accretion of solids is mainly in the form of \textit{planetesimals} (0.1 - 100 km size objects), there is an intermediate stage of growth in which the region of gravitational influence of the protoplanet is depleted with solids, but the accretion luminosity is enough to sustain the growth of the envelope at a small rate \citep[the so-called \textit{phase-2} of][]{P96}. In this scenario, phase-2 is usually the longest (of the order of 10$^6$ yrs), and therefore, the mass the planet has during this period plays a crucial role in the possibility of forming a mini-Neptune. 
In the classical core accretion model of \citet{P96}, the formation timescale was of the order of several million years. Following studies have shown that the long formation timescale problem can be overcome including various processes/assumptions such as planetary migration \citep{Alibert05}, specific conditions of planetesimal sizes \citep{Fortier13}, envelope's opacity \citep{Movshovitz10} and the replenishment of scattered planetesimals from the embryo's feeding zone \citep{TanakaIda99}.  

An alternative scenario that overcomes the long formation timescale in the core accretion model rises from considering the accretion of \textit{pebbles}, with typical sizes of $\sim 10$-cm  \citep{OrmelK10, Lambrechts12,Lambrechts14b,Bitsch15}. Since pebbles move at Keplerian velocities and are small, they experience a strong headwind from the sub-Keplerian gas, which leads to orbital decay. 
This leads to a flux of pebbles that can be accreted efficiently by the protoplanet due to gas friction that significantly slows them down \citep{Lambrechts12}.
The accretion rate of solids in the pebble accretion case is typically high ($\sim 10^{-6}-10^{-5} \, \Mearth/$yr) for low-turbulent disks during the entire growth \citep{Morby15}. This stops  when the protoplanet perturbs the disks' structure, producing pressure bumps that hinder pebble accretion when the protoplanet has a mass of $\sim 20$ \ME \, \citep{Lambrechts14}, higher than the masses we focus on in this work. 

Many studies have focused on the formation of mini-Neptunes in-situ, given that, due to observational biases, most of the exoplanets with radii between 1 and 4 \RE \, have periods of less than 50 days \citep{DressChar15}. All studies of this type concluded that gas accretion inside the iceline during the disk's lifetime is very rare \citep{IkomaHori12, Lee16,Inamdar15,BodLiss14}. Still, even beyond the iceline, the acquisition of a gaseous envelope of 10 \% of the planet mass or higher requires very specific disk conditions \citep{Lee16, Rogers11}.

The formation of low-mass, gas-rich planets is difficult to understand in the context of the classical core accretion model where all the solids are assumed to sink to the core because in that case protoplanets begin to accrete considerable amounts of gas only when the core reaches a critical mass of $\sim 10 $ \ME \, \citep{mizuno80, P96}. 
In this paper we include the effect of envelope enrichment by the sublimation of icy planetesimals/pebbles, whose effect was initially studied by \citet{Podolak88, Iaroslav07} by computing the deposition of solids in the primordial H-He envelopes. Later on, this process was incorporated by \citet{HI11, Lambrechts14, Venturini15} in the calculation of critical core masses, and fully implemented in planetary growth in \citet{Venturini16}. We follow the work of \citet{Venturini16} which showed that envelope enrichment can lead to the accretion of 
large amounts of H-He  before the onset of rapid gas accretion. We concentrate on the key physical properties that can change the efficiency of mini-Neptune formation, including the solid accretion rate, formation location, and atmospheric opacity. The planets are assumed to form beyond the iceline and migrate inwards afterwards. 
We find that under a quite broad parameter space, the formation of mini-Neptunes is likely to occur, especially when envelope enrichment is considered.

\section{Methods}\label{method}
The planetary formation is modelled by solving the structure equations with global energy conservation as used in standard core accretion calculations (see Helled et al. 2014 for review). 
Unlike previous studies, we account self-consistently for the envelope enrichment by icy planetesimals/pebbles and its effect on the planetary growth.  
The simulations begin with an embryo of 0.01 \ME, and the pollution of the envelope with volatiles starts once the envelope is thick enough for disrupting the accreted solids (see Sect.\ref{Mdotz} for details). 

Once enrichment begins, the water (H$_2$O) is assumed to remain homogeneously mixed in the envelope (i.e., uniform envelope metallicity). In order to simplify the calculation, and since a large fraction of the rocky material (SiO$_2$) tends to reach deeper regions compared to volatiles \citep{Iaroslav07,Lozovsky17} (although the exact fraction is not well-determined) we assume that all the refractory material is added the core.

For the equation of state of the envelope, we use a mixture of H-He \citep{SCVH} with H$_2$O \citep[ANEOS,][]{aneos} while for the core we implement the mass-radius relation of \citet{Valencia10} for rocky material, which accounts for core compressibility. The opacity is calculated by considering both gas and dust opacities and is explained in Sect.\ref{sect_opac}. More details on the formation model can be found in \citet{Venturini16}.

\subsection{Definitions of mini-Neptunes, Neptunes, and Occurrence Rate}\label{defs_mN}
In this work, we use the term {\it Neptunes} to describe planets with masses between 10 and 20 M$_{\oplus}$ and a H-He mass fraction of 0.1-0.25 \citep{Helled11, Nettelmann13a}. We define a \textit{mini-Neptune} with the same mass fraction of H-He (denoted as $\fhhe$), but for a total planet mass in the range of $0 < M_P \leqslant 10 $ \ME.  
We define $t_{\text{mini-Nept}}$ as the time spent by the protoplanet in the mini-Neptune region (i.e., with  $ 0.1  \leqslant \fhhe  \leqslant  0.25$ and $0 < M_P  \leqslant 10 $ \ME ),

We determine the likelihood to obtain a mini-Neptune under different formation conditions, i.e., the \textit{mini-Neptune occurrence rate}. This definition should capture the fact that if the core becomes critical while the disk is still present, a significant mass of gas is accreted and the planet will become a gas giant, not a mini-Neptune. In order to form a mini-Neptune, the growing protoplanet must spend a significant time in the mini-Neptune-regime until the gaseous disk disappears.

As a result, our definition of the occurrence rate should include a certain prior of the disk lifetime.  
Observations show that disks have lifetimes of a few million years, and that they should not disappear in the first million years, regardless of the stellar mass. Then, at times 1-3 Myr a sharp decrease in disk frequency occurs, and finally, no disks are observed at ages around $\sim 10$ Myr \citep{Ribas15}. The fraction of stars with disks at different ages has been fit with an exponential decay of 2.5 Myr \citep{Mamajek09} or 4 Myr \citep{Pfalzner14}. Therefore, the cumulative distribution of the disk's lifetime is $cdf = 1-\text{exp}(-time/\tdisk)$, but flat during the first $\sim$1 Myr if one accounts for the recent observations by \citet{Ribas15}. Consequently, we adopt the following probability density function of disk lifetime:

\begin{equation}
p(t) = \frac{1}{\tdisk }\frac{e^{-t/\tdisk}}{(e^{-t_0/\tdisk} -e^{-t_M/\tdisk})}
\end{equation}
if $ t_0 \le t\le t_M$ and p(t) = 0 otherwise.  $t_0$ is the minimum time at which the disk can disappear and $t_M$ the maximum, and they are set to be 1 Myr and 10 Myr, respectively. Note that the integral between $t_0$ and $t_M$ yields 1, as it must for any probability density function.
The mini-Neptune occurrence rate is therefore defined as the integral of $p(t)$ in the domain where  $t_{\text{mini-Nept}} \ne 0$, which yields:

\begin{equation}\label{eq_occ_rate}
f_{MN} = \frac{e^{-t_{i}/\tdisk} - e^{-t_f/\tdisk}}{e^{-t_0/\tdisk} -e^{-t_M/\tdisk}}
\end{equation}
where  $t_i$ is the time when the planet enters the mini-Neptune region if $t_i \geq t_0$ or $t_0$ otherwise, while $t_f$ is the time when the planet exits the mini-Neptune region if $t_f \leq t_M$ or $t_M$ otherwise.

It should be clear, however, that our calculated "occurrence rates" are given as a guidance for the likelihood to form the planets under given theoretical conditions.

Since low-mass, low-density exoplanets could have a wider range of H-He mass fractions, we also consider a second, less restricted definition of a mini-Neptune, assuming the same mass range, but allowing the mass fraction of H-He to be in the range of $ 0.05  \leqslant \fhhe \leqslant 0.5$. As we show below, this definition (to which we refer as {\it extended}) yields occurrence rates typically higher by a factor of 2 with respect to the first, {\it restricted} definition.
\newpage
\subsection{Disk Model}
Our baseline disk model is the standard minimum mass solar nebula (MMSN) of \citet{Weiden77} and \citet{Hayashi81}, where the initial surface density of solids ${\Sigma}_s$ is given by the power-law: 

${\Sigma}_s =  {\Sigma}_0 \, Z_0 \, (r/\text{AU})^{-p} $
with $Z_0$ = 0.018 (the disk metallicity), $p = 3/2$,  and ${\Sigma}_0$ = 1700 \gcmcuad. A comparison with other disk parameters is discussed in Sect.{\ref{other_params}}.

We consider in-situ formation between 5 and 30 AU. 
For the temperature profile we adopt a standard passive disk \citep[see, e.g.,][]{Armitage10,Guilera11}, with $T(r) = T_0 \, (r/AU)^{-1/2}$ and $T_0=280$ K. The protoplanet is assumed to be embedded in the gaseous disk during its growth, thus the outer temperature of the envelope matches the disk temperature at the position of the protoplanet.

For the disk scale height we assume a vertically isothermal disk. Using this, the definition of the disk scale height ($H_{\text{gas}} = c_s/\Omega$) and the temperature profile described above, it can be shown that the aspect ratio ($h = H/r$) is given by $h(r) = h_0  (r/AU)^{1/4}$, with $h_0 = \sqrt{(k_B \, AU/ \mu m_H G M_*)T_0} $, with $h_0 \simeq 0.036$ for the chosen temperature profile, and the remaining symbols are in standard notation.

Besides temperature, the other necessary boundary condition to integrate the structure equations for the planetary envelope is the outer pressure. To obtain this we assume a classical power-law profile: $P(r) = P_0  (r/AU)^{-q}.$
The exponent $q$ is derived from the ideal gas law and the condition that the column gas density is: $\rho_\text{gas} = \Sigma_\text{gas} / H_\text{gas}$. We get $q = p + 7/4$ and $P_0 = \sqrt{(k_B \, GM_*/ \mu m_H  AU^3)T_0 \Sigma_0}$, which for a MMSN disk is $P_0  \simeq  36.5$ dyn/cm$^2$.

\subsection{Accretion rate of solids}\label{Mdotz}
In reality, the solids in the disk should have a size distribution, ranging from mm-dust to hundred-km sized planetesimals. This size distribution evolves as pebbles grow and convert into planetesimals \citep{Drazkowska16}, and as planetesimals fragment \citep{Guilera14}. However, since the size distribution of solids and its time evolution is still poorly known, we consider a given size and concentrate on the effect of envelope enrichment assuming the accretion of \textit{just pebbles} and \textit{just planetesimals}, and compare the two. In the case of pebbles, we assume that envelope enrichment begins when the envelope mass is $\Menv = 10^{-6} \, \Mearth$, whereas in the case of planetesimal enrichment, when $\Menv = 2 \times 10^{-4} \, \Mearth$ \citep[see Table 4 of][]{Venturini16}.

\subsubsection{Pebble accretion}
For pebble accretion we use the prescription derived initially by \citet{Lambrechts14b} (hereafter LJ14), and adapted later to include the effect of disk turbulence \citep{Morby15, Brasser17}. In this model, pebbles with Stokes number smaller than $\sim$ 1 drift towards the star and are intercepted by the growing embryo. The surface density of gas decays exponentially with time to mimic gas dissipation and is given by:

\begin{equation} 
\Sigma_\text{gas} =  \Sigma_0 \, (r /\text{AU})^{-p} e^{-t/\tdisk} \ ,
\end{equation}
where $\tdisk$ is the disk's lifetime, which in our baseline model is taken as $\tdisk = 3$ Myr.
The initial surface density of dust is simply $ {\Sigma}_s =  Z_0 \, \Sigma_\text{gas}$. The surface density of pebbles ($\Sigma_P$) can be found from the definition of the pebble flux: $\dot{M}_F = 2 \pi r v_r \Sigma_P$. The pebble flux can be computed, on the other hand, from the condition of equating the growth timescale with the radial drift timescale (see LJ14 for the details).
The efficiency of pebble accretion depends on the pebble scale height: $H_{\text{peb}} = H_{\text{gas}} \sqrt{\alpha / \tau_{\text{f}}}$,
where $\alpha$ is the viscosity-parameter \citep{SS73} and $\tau_{\text{f}}$ is the Stokes number of the drifting pebbles, which can be computed assuming a dominant particle size as described in the simple dust growth model of \citet{Birnstiel12} and LJ14.

If the Hill radius of the planet exceeds the pebble scale height, pebble accretion is 2D and is given by:
\begin{equation}
\dot{M}_{z, 2D} = 2 \,\bigg( \frac{ {\tau}_f }{0.1} \bigg)^{2/3} R_H v_H {\Sigma}_P \ ,
\end{equation}
where $R_H$ is the planet's Hill radius, $v_H$ the Keplerian velocity at a distance of the Hill radius from the center of the planet, and  ${\Sigma}_P$ the surface density of pebbles at the position of the planet.

The accretion rate of pebbles transitions to the less efficient 3D fashion if:
\begin{equation} \label{peb_3D}
\frac{\pi (\tau_{\text{f}}/0.1)^{1/3} R_H}{ 2\sqrt{2 \pi}} < H_{\text{peb}}  \ .
\end{equation}
In this case the accretion rate is given by \citep{Brasser17}:
\begin{equation}
\dot{M}_{z, 3D} = \dot{M}_{z, 2D} \bigg( \frac{\pi (\tau_{\text{f}}/0.1)^{1/3} R_H}{ 2\sqrt{2 \pi} H_{\text{peb}} } \bigg) \ .
\end{equation}
Therefore, when pebble accretion becomes 3D, its value depends on $\alpha$, decreasing for higher disk's viscosity. If the disk is turbulent ($\alpha>10^{-3}$), the growth of the planet is typically so slow that gas accretion within the disk's lifetime is insignificant, and hence, mini-Neptune formation unlikely, as we show in Sect.\ref{other_params}.

\subsection{Planetesimal accretion}
For planetesimal accretion we use the high accretion rates given by \citet{Rafikov11} . 
This rate is appropriate for small planetesimals ($\sim$ 100 m - 1 km) for a dynamically cold planetesimal disk \citep[see][for details]{Goldreich04}.
\begin{equation}\label{rafikov}
\dot{M}_z = 6.47 \Omega P^{1/2} \Sigma_s R_H^2 ,
\end{equation}
where $\Omega$ is the Keplerian frequency of the protoplanet,  and $P = R_c / R_H$, $R_c$ being the capture radius of the planet \citep[computed  following the prescription of][]{Inaba03}.  
Large planetesimals cannot be accreted efficiently since they are less affected by gas drag when passing through the planetary envelope \citep[e.g.,][]{P96, HelledBod14,Inaba03}.  
In addition, planetesimals' eccentricities and inclinations tend to increase during the planetary growth due to the perturbation of the embryo \citep{IdaMak93,Ormel10}, which increases the relative velocities of planetesimals, reducing the accretion efficiency. 
Small planetesimals ($\sim100$ m) can still be accreted due to the effect of gas drag, allowing for the formation of gas-rich objects before disk dispersal \citep{Fortier13}. 
We therefore focus on planetesimal accretion rates that correspond to relatively small sizes. Nevertheless, we discuss the possibility of lower accretion rate of solids (i.e., larger planetesimals) in Sect.\ref{other_params}.

\subsection{Envelope's Opacity}\label{sect_opac}
The assumed opacity of the planetary envelope has a crucial role in determining the growth rate of the planet \citep{Podolak03, Movshovitz10}. 
For our baseline model we assume gas opacities from \citet{Freedman14}, which account  for the increase of gas opacity caused by the raise in metallicity. For the dust opacities we adopt the analytical model of \citet{Mordasini14}, which includes dust growth and settling. Minimum values of these dust opacities in the outer layers of the envelope are rather low ($\sim$ 0.001-0.01 cm$^2$/g), hence we refer to this case as {\it low dust opacity}. Although these low opacity values are justified as well by other works \citep[e.g.,][]{Movshovitz08, Ormel14}, higher dust opacities could be a result of the ongoing ablation of accreted planetesimals/pebbles. Indeed, the dust opacities of \citet{Mordasini14} were computed assuming a unique dominant size of grains. If ablation of solids is dominant in the upper atmosphere, most of the mass of solids could reside in smaller dust aggregates, and hence the opacity could be higher. Also, the recondensation of upstreaming water vapor (cloud formation) could increase the opacities. To account for this possibility we also run calculations with dust opacities increased by a factor of 100 and we refer to this as the {\it high dust opacity} model.

 \section{Results} 
Below we present the formation histories of the planets and the derived occurrence rates accounting for the different possible accretion rates, envelope enrichment, opacities, and formation locations. Sections \ref{pebbles_specific} and \ref{planetesimals_specific} show results for our baseline MMSN disk model. Further sections explore the possibility of other disk parameters.

\subsection{Pebble Accretion}\label{pebbles_specific}

For the case of pebble accretion with low dust opacities, we find that mini-Neptunes are more likely to form when envelope enrichment is included.  
Figure \ref{bulk_comp} shows that if envelope enrichment is not included, the formation path of the planet crosses the parameter space of mini-Neptunes if the embryo grows at 20 AU but for times longer than the disk's mean lifetime, yielding a mini-Neptune occurrence rate of 9\% and 19\% for the \textit{restricted} and \textit{extended} cases, respectively. At $a= 5$ AU without enrichment and low opacities, the formation path only crosses the mini-Neptune regime by invoking the \textit{extended} definition. In other words, for this case, only mini-Neptunes with $\fhhe$ < 0.1 can be formed, but with a negligible occurrence rate (1\%).
When envelope enrichment is considered, the protoplanet crosses the mini-Neptune region, both at 5 and 20 AU. However, at $a= 5$ AU the crossover mass is reached so early ($\tcross$ = 0.65 Myr), that the planet becomes a gas giant in $\sim$1 Myr, giving a mini-Neptune occurrence rate of zero. At $a = 20$ AU the occurrence rate is 27\% with the  \textit{restricted} definition and 41\% with the \textit{extended} definition.

The fact that mini-Neptunes appear more easily when envelope enrichment is included, can be understand as follows. When the volatiles are allowed to mix in the primordial H-He envelopes, the mean molecular weight increases, which translates into an increase in the envelope's density and self-gravity. The envelope is therefore more prone to contract, allowing accretion of  H-He in larger amounts once enrichment begins.  
This leads to the formation of low-mass planets with larger fractions of H-He than in the case without envelope enrichment, and in addition, to relatively short formation timescales. 
It is important to note that for the case of pebble accretion analyzed in this section, $\alpha$ is set to 10$^{-5}$ (laminar disk), which corresponds to high solid accretion rates.  Hence, the timescale to form a gas giant is short, as summarized in Table \ref{tab1}. Results with other values of $\alpha$ are shown in Sect.\ref{other_params}.

When higher dust opacities are invoked for the envelope, because of the slower cooling, formation timescales are longer (see Table \ref{tab2}). In principle, this implies longer times spent in the mini-Neptune region, and therefore, higher occurrence rates. The maximum occurrence rate of mini-Neptunes is found to be $\sim$79\% (\textit{extended} definition) for the case with envelope enrichment, and $a = 20$ AU. Without envelope enrichment, mini-Neptunes formation is very unlikely. Even when choosing different disk parameters (see Sect.\ref{other_params}), we find that typically envelope enrichment is required for the formation of mini-Neptunes with pebble accretion. The results for pebble accretion with low and high opacities are summerized in Table \ref{tab1} and Table \ref{tab2}, respectively. 
Since pebbles are small objects they are likely to enrich the envelope and to increase the opacity, and we expect the high occurrence rates for pebble accretion to be more appropriate.

\subsection{Planetesimal Accretion}\label{planetesimals_specific}
The case of planetesimal accretion is rather different. As we discuss above, the classical in-situ growth of a planet by planetesimal accretion is characterized by three phases of growth \citep{P96}. 
During phase-2, the planetary mass increases very slowly, and the heavy-element mass at the beginning of this stage (known as the isolation mass, $\Miso$) strongly depends on the amount of solids in the embryo's vicinity. 
In other words, for more massive and/or metal-rich disks, the higher  $\Miso$ is. 
Indeed, \citet{P96} showed that  $\Miso \sim a^3 \Sigma_s^{3/2} $, with the exact relation being:

\begin{equation}
\Miso = \mathbf{C} \, ( {\Sigma}_0 \, Z_0 )^{3/2} \, \bigg( \frac{a}{\text{AU}}\bigg)^{\frac{3}{2} (2-p)}, 
\end{equation}
with  $\mathbf{C} \approx 0.0026$ to obtain $\Miso$ in Earth masses.
  
For a MMSN disk profile $\Miso \propto  a^{3/4}$. The dependence of $\Miso$ on the semi-major axis results in $\Miso \approx 1.5 \, \Mearth$ at $a = 5$ AU and $\Miso \approx 4.2 \, \Mearth$ at  $a = 20$ AU (see Fig.\ref{planet_growth}). Due to the relatively long timescale of phase-2, the value of $\Miso$ plays a crucial role on determining whether a protoplanet can be in the mini-Neptune region. If the combination of $a$ and $\Sigma_s$ results in a $\Miso < 10 \, \Mearth$, then the protoplanet is likely to spend some time in the mini-Neptune region (see Sect.\ref{pla_general}).

\begin{figure}
\begin{center}
\includegraphics[width=0.9\columnwidth]{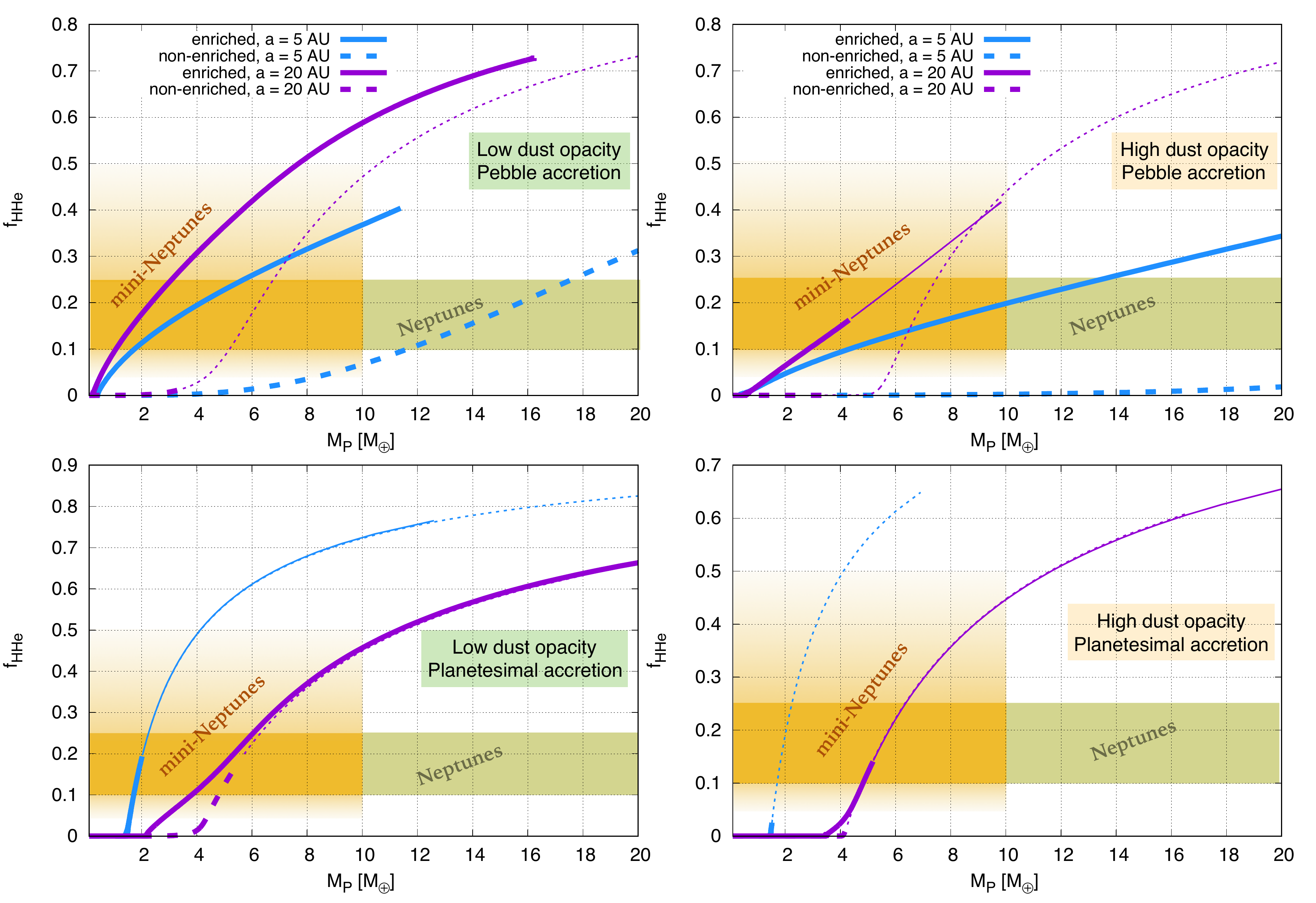}
\caption{Calculated planet formation paths. Shown is the H-He mass fraction vs.~planetary mass ($\Mp$) at different locations, for pebble and planetesimal accretion and different opacities. 
The solid and dashed curves correspond to \textit{enriched} and \textit{non-enriched} cases, respectively. The thick parts of the curves indicate the time within the disk's mean lifetime. The green areas represent the "Neptune-regime", while the orange ones the "mini-Neptunes-regime" (dark orange - \textit{restricted}; light orange - \textit{extended}). We find that without envelope enrichment, the formation of mini-Neptunes is rare in the pebble accretion scenario. For the case of planetesimals and high dust opacities, formation timescales are generally too long for the protoplanet to enter in the mini-Neptune region within the disk's mean lifetime of 3 Myr (see text for details).}
\label{bulk_comp}
\end{center}
\end{figure}

\begin{figure}
\begin{center}
	\includegraphics[width=0.9\columnwidth]{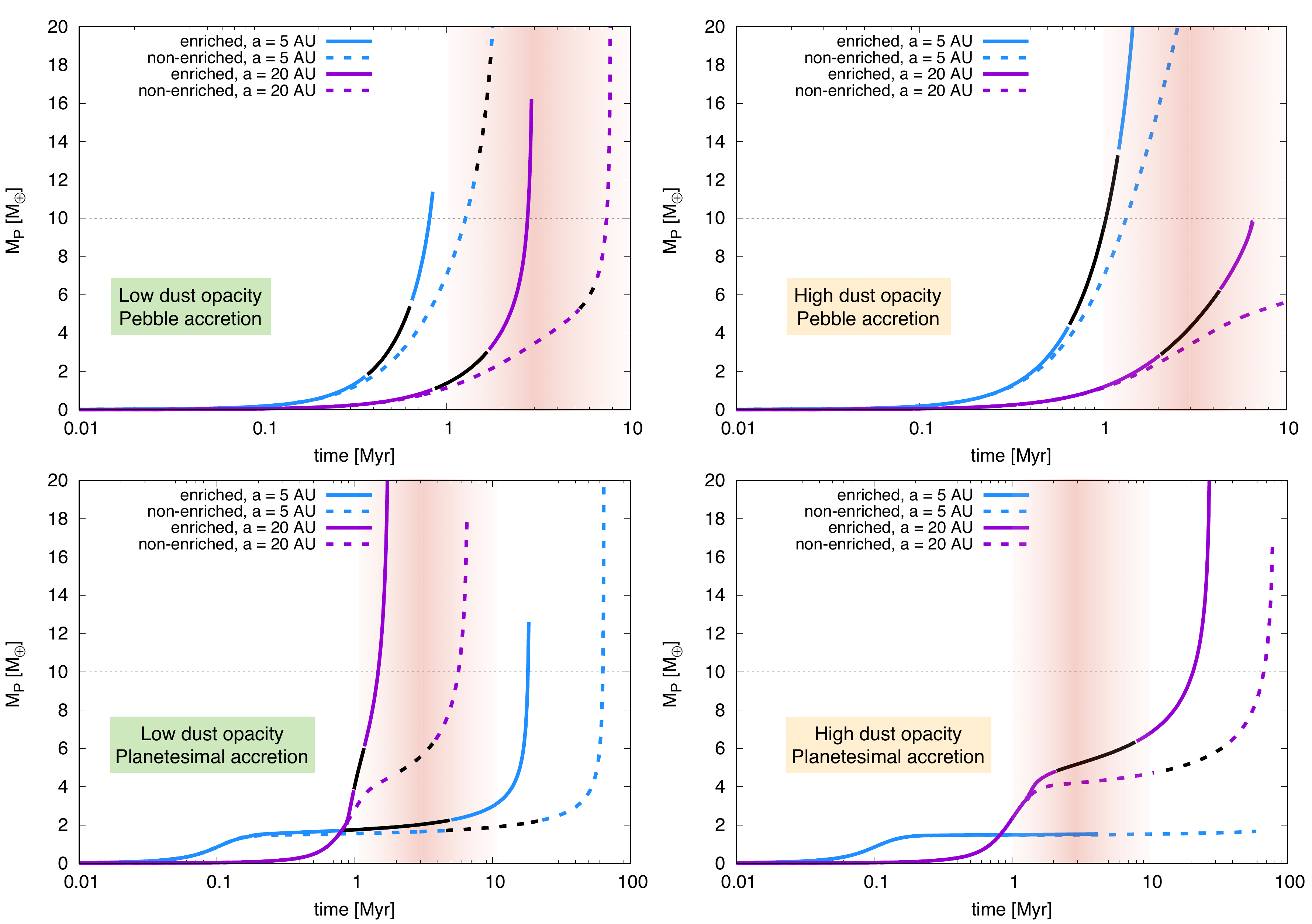}
	\caption{Planetary growth for the same simulations shown in Figure \ref{bulk_comp}. The solid black parts indicate the time during which the protoplanets have  $0.1\leqslant \fhhe  \leqslant 0.25$. 
	The color gradient in the shaded area corresponds to the expected ages of disk dispersal with a mean disk lifetime of 3 Myr (see Sect.\ref{defs_mN}). }
\label{planet_growth}
\end{center}
\end{figure}

As shown in Fig.\ref{bulk_comp} and Table\ref{tab1}, with low dust opacities, at $a = 5$ AU the occurrence rate of mini-Neptunes is extremely high when envelope enrichment is included (83\% with the \textit{restricted} definition and 100\% with the \textit{extended} one), and lower without envelope enrichment (33\% and 76\%). At 20 AU with envelope enrichment, the crossover time is shorter than at 5 AU ($\sim$ 1.5 Myr), and therefore the mini-Neptune occurrece rates are a lower (see Fig.\ref{planet_growth} and Table \ref{tab1}).

When we use high dust opacities, $\Miso$  is unchanged but the formation timescales are so long --particularly when neglecting envelope enrichment-- that although the planetary mass remains low during the first 3 million years of growth, the gas accretion rate is negligible. 
As a result, the H-He mass fraction is small in these  cases, typically being less than 5\% for $t \leq 3$ Myr (see Fig.\ref{bulk_comp}), and mini-Neptunes are unlikely to form, unless envelope enrichment is invoked at large semimajor-axes. For instance, for the enriched case at $a =20$ AU, the occurrence rate is found to be 66\% with the \textit{restricted} definition and 82\% with the \textit{extended} one.

In the case of planetesimal accretion, the results are very sensitive to the assumed density profile in the disk, due to the dependence on $\Miso$. It should be noted that $\Miso$ is independent of $a$ only for a disk with $p=2$, which was the value chosen in \citet{P96}. Observations typically infer values of $p \sim 1$ \citep{Andrews10}. For planetesimal accretion, the flatter the disk, the stronger the dependence of $\Miso$ on the location of the embryo. 

One difference between pebble  and planetesimal accretion is that with pebbles the occurrence rate of mini-Neptunes is always higher when envelope enrichment is considered. With planetesimals, it is more difficult to draw a general conclusion. If phase-2 is so long that the mini-Neptune region is crossed at times larger than the disk's lifetime, then envelope enrichment favors the occurrence of mini-Neptunes, because it shortens phase-2 and allows the protoplanet to cross the mini-Neptune region within the disk's lifetime (see, e.g, the enriched and non-enriched cases at $a=5$ AU of Fig.\ref{planet_growth}). 
On the other hand, if the duration of phase-2 is of a few Myr when the envelope is pure H-He, envelope enrichment leads to formation timescales shorter than the disk's mean lifetime, promoting the formation of gas giants (simulations at $a = 20$ AU and low-dust opacities, see Fig.\ref{planet_growth}).

\begin{figure}
\begin{center}
\includegraphics[width=0.9\columnwidth]{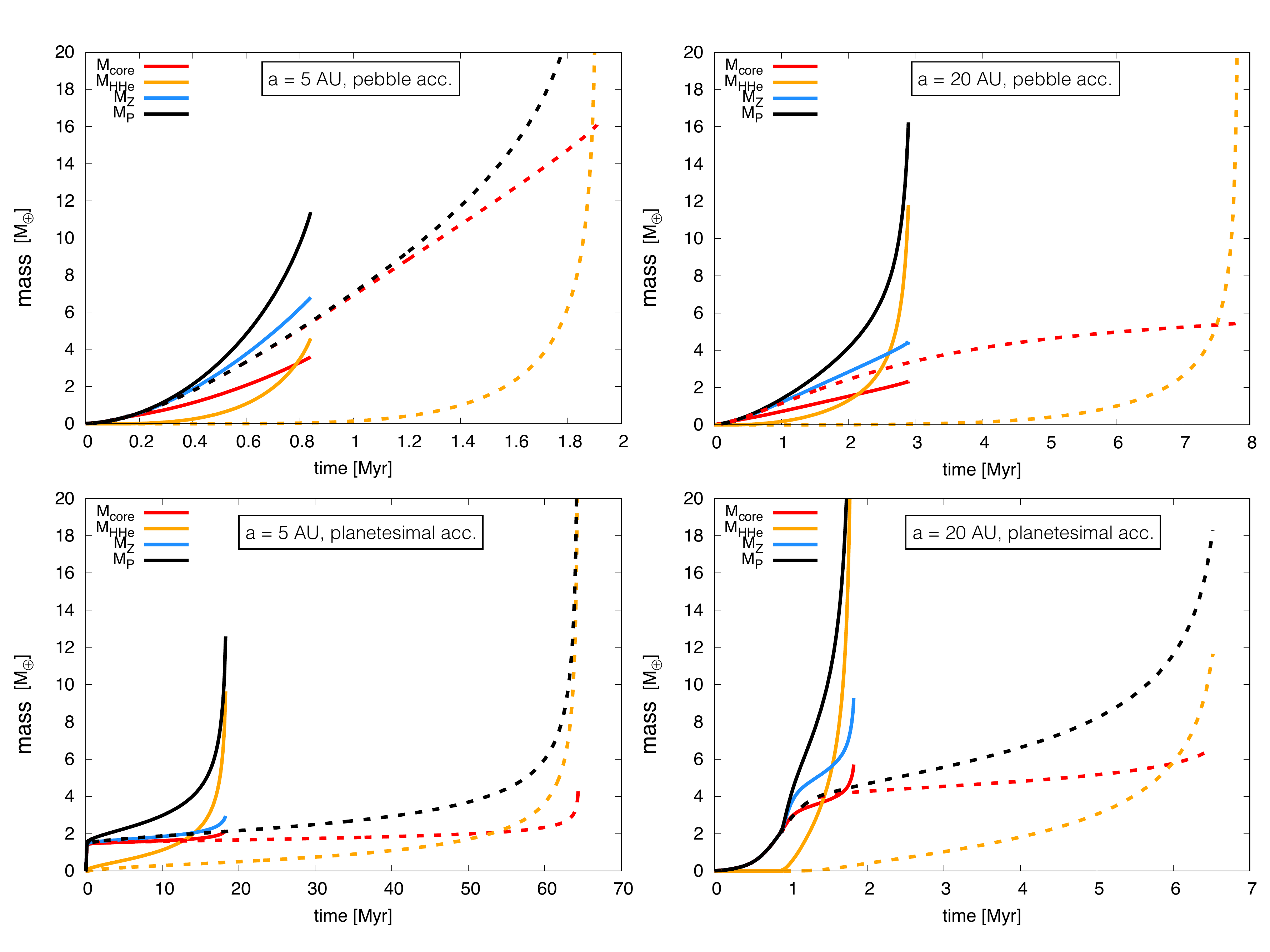} 
\caption{The planetary growth history at 5 and 20 AU for the low-dust opacity cases of Fig.\ref{planet_growth}. Shown are the mass of the core ($\Mcore$), total mass of heavy elements (M$_Z$), the mass of H-He ($\MHHe$) and the total planetary mass ($\Mp$) as a function of time. The solid and dashed lines correspond to the enriched and non-enriched cases, respectively. }
\label{App1}
\end{center}
\end{figure}

\begin{figure}
\begin{center}
\includegraphics[width=0.9\columnwidth]{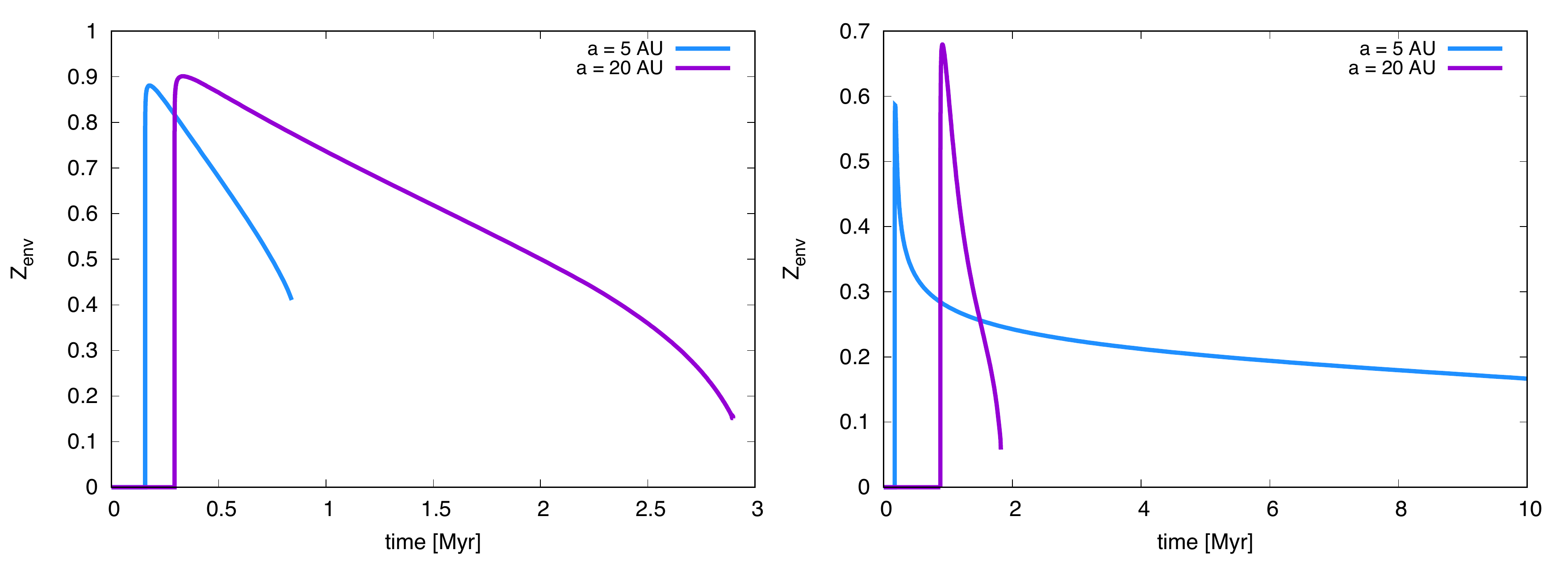} 
\caption{Evolution of the envelope's metallicity for the enriched cases of Fig.\ref{App1} for pebble accretion (left) and planetesimal accretion (right) assuming low-dust opacity. Once envelope enrichment begins, the volatile material (water) is assumed to remain in the envelope and mix with the H-He while the rocks are deposited into the core.}
\label{App2}
\end{center}
\end{figure}

Examples of the growth histories of the planets are shown in Fig.\ref{App1}. This figure shows the mass of the core ($\Mcore$), total mass of heavy elements (M$_Z$), the mass of H-He ($\MHHe$) and the total planetary mass ($\Mp$) as a function of time. Shown are the cases for pebbles (top) and planetesimal (bottom) accretion for the two different formation locations at 5 AU (left) and 20 AU (right) for the low dust opacity case. 
Figure \ref{App2} shows the evolution of the envelope metallicity for the enriched cases. Since the envelope is thin at early times, the metallicity increases very rapidly when enrichment begins. Further details on the model can be found in \citet{Venturini16}.

\subsection{"Sweet spots" for mini-Neptune formation: testing other disk parameters}\label{other_params} 
In this section we investigate the sensitivity of the results to the assumed disk parameters ($\alpha$, $\Sigma_0, Z_0, p$). For the simulations presented in this section we used the baseline low-dust opacities.

\subsubsection{Pebbles}\label{peb_general}
\begin{figure}
\begin{center}
	\includegraphics[width=0.8\columnwidth]{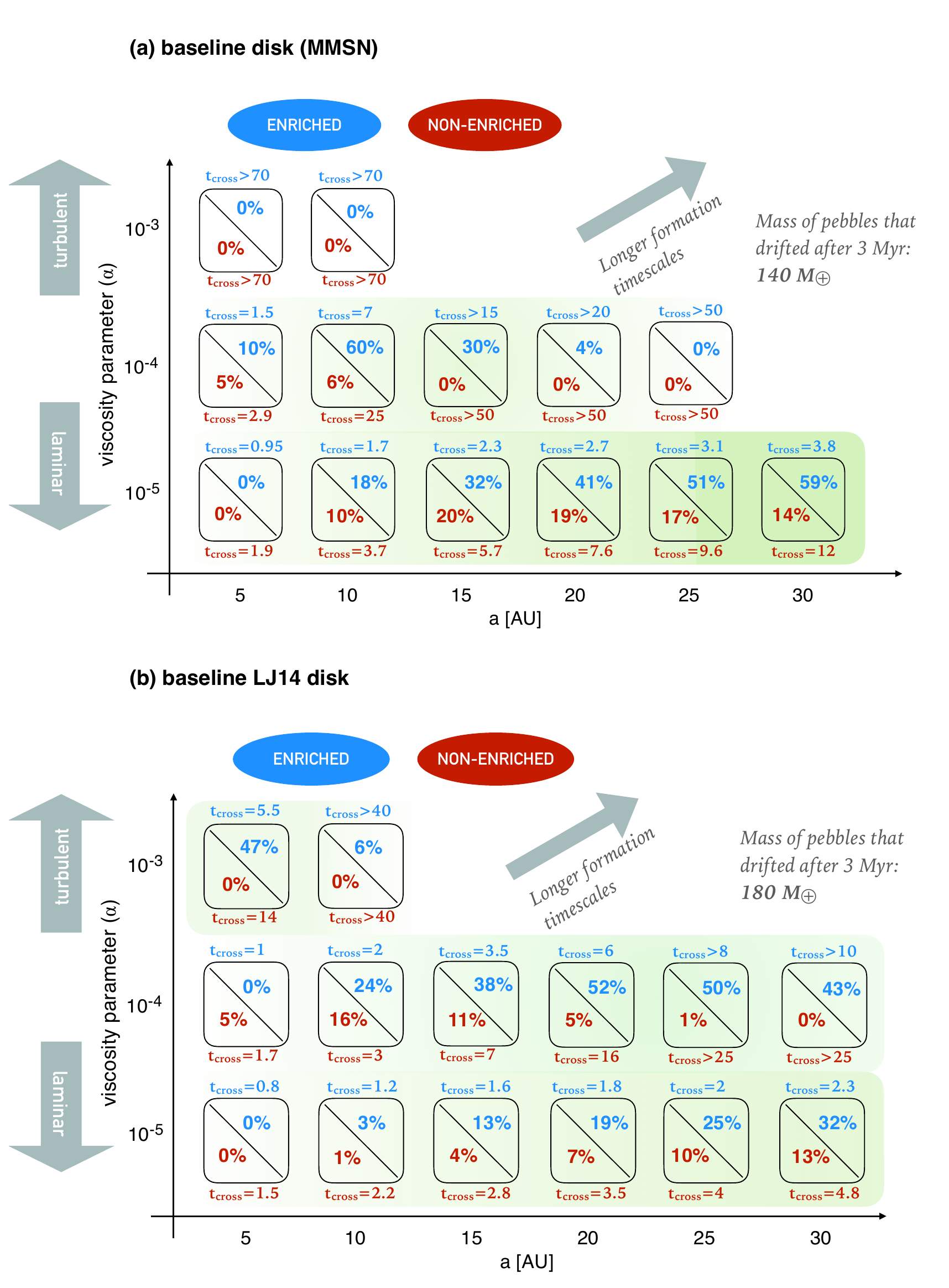}
	\caption{Mini-Neptune occurrence rate  (\textit{extended} definition) vs.~semimajor axis ($a$) and the viscosity parameter ($\alpha$). Shown are the \textit{enriched} (light-blue) and \textit{non-enriched} (red) cases. We also indicate the crossover time in Myr for the corresponding simulation. Top: (a) MMSN disk (p =1.5, $\Sigma_0 = 1700$ \gcmcuad, $Z_0 = 0.018$). Bottom: (b) Baseline disk of LJ14 (p = 1, $\Sigma_0 = 500$ \gcmcuad, $Z_0 = 0.01$). The green-shaded area highlights the cases where the formation of mini-Neptunes is possible. }
\label{pebbles_baseline_params}
\end{center}
\end{figure}

The growth of a planet via pebble accretion depends sensitively on the disk turbulence \citep[][submitted]{Ormel17}. Certainly, Fig.\ref{pebbles_baseline_params}a shows that for our baseline MMSN disk, the occurrence rate of mini-Neptunes depends drastically on the value of $\alpha$. The more turbulent the disk is, the more spread the pebbles are vertically, and therefore, the more difficult it is for the protoplanet to intercept them. For instance, for our basline MMSN disk (Fig.\ref{pebbles_baseline_params}a)  at $a = 10$ AU and $\alpha = 10^{-3}$, the core can just grow from 0.01 \ME \, to 0.05 \ME \, during the disk's mean lifetime. This happens because for this case all the growth occurs in the 3D regime (Eq. \ref{peb_3D}), and for these particular disk's parameters the accretion rate of pebbles is of the order of $10^{-8}$ \ME/yr. On the contrary, for low values of $\alpha$ (i.e., $10^{-5}$), the accretion rate is in the 2D form, and hence, typically high ($\sim 10^{-5}-10^{-6}$ \ME/yr). This is clear from the low values of crossover times  shown in Fig.\ref{pebbles_baseline_params}, especially for $5 < a < 15$ AU. 
 
The most striking result from Fig.\ref{pebbles_baseline_params} is that the formation of mini-Neptunes is much more likely when envelope enrichment is considered, similarly to the results we get in Sect.\ref{pebbles_specific}. Not only that the inferred occurrence rate is higher when envelope enrichment is included, but also the parameter space is much broader.

The detail analysis of Fig.\ref{pebbles_baseline_params} is not so intuitive, but the general trend is that if formation timescales are too short, the occurrence rate of mini-Neptunes is low and the most likely output is the formation of a gas giant. Also, if formation timescales are too long (for instance, the cases with $\tcross > 10$ Myr), the protoplanet accretes negligible amounts of H-He during the disk's lifetime. Thus, the formation of mini-Neptunes is unlikely, the most typical output in these cases are lunar- to Earth-mass embryos. 

We find that, besides the disk viscosity, the crucial variable that dictates the likelihood to obtain a mini-Neptune is the amount of drifting pebbles during the disk's mean lifetime. For the simulations shown in the two panels of Fig.\ref{pebbles_baseline_params}, the amount of pebbles that drifted after 3 Myr is 140 $\Mearth$ for the MMSN disk and 180 $\Mearth$ for the baseline LJ14 disk ($p = 1$, $\Sigma_0$ = 500 \gcmcuad , $Z_0 = 0.01$). For both disk models the formation of mini-Neptunes is likely, and occurs in a broad parameter space, as long as envelope enrichment is included. 

\begin{figure}
\begin{center}
	\includegraphics[width=0.8\columnwidth]{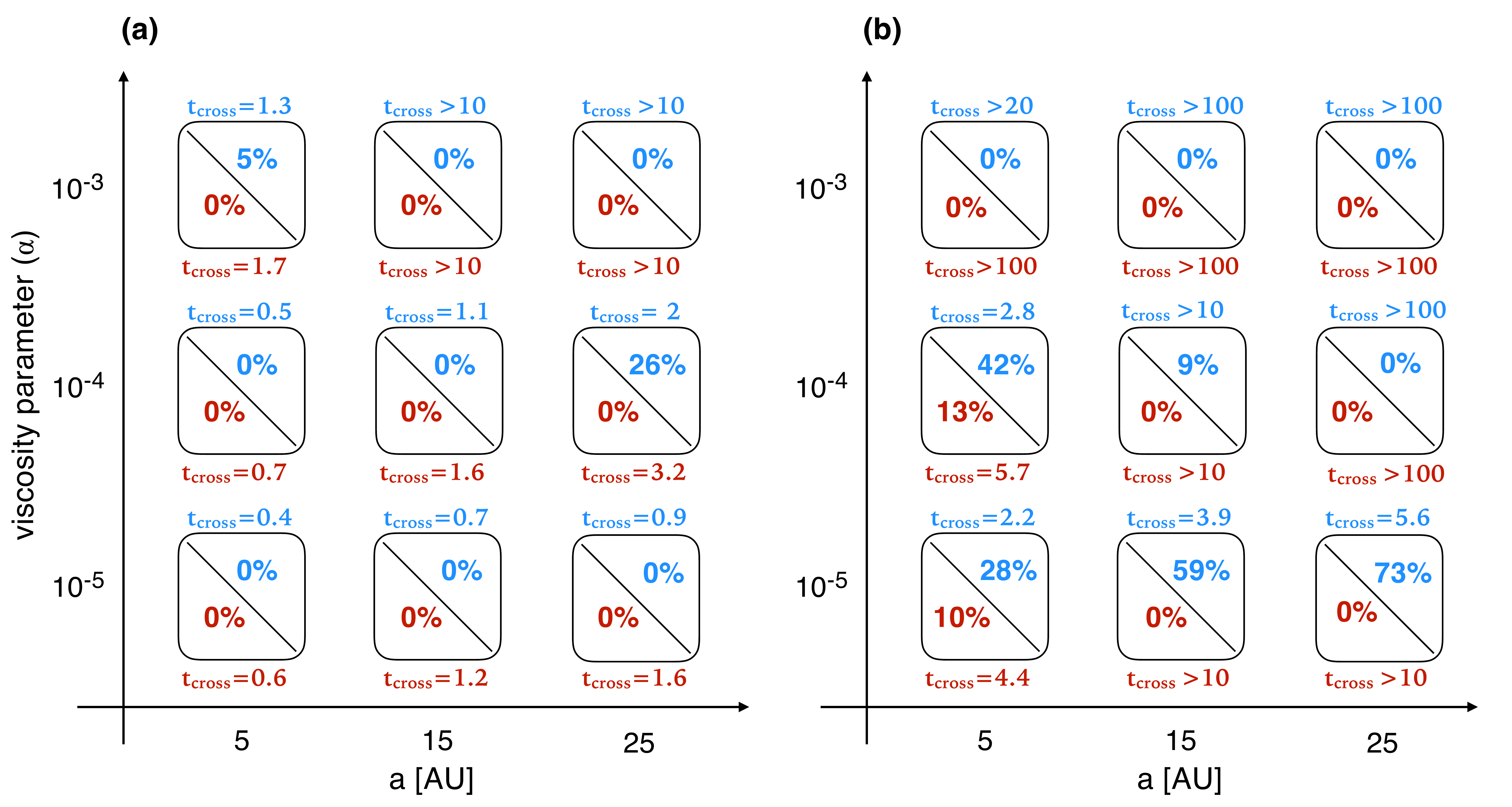}
	\caption{Mini-Neptune occurrence rate  (\textit{extended} definition) vs.~semimajor axis ($a$) and the viscosity parameter ($\alpha$) for disks with same profile and metallicity as the baseline of LJ14 ($p=1$, $Z_0 = 0.01$), but different total mass. 
	Left: (a) $\Sigma_0 = 1000$ \gcmcuad, with the total mass of pebbles that drifted after 3 Myr being 365 \ME \,. Right: (b) $\Sigma_0 = 250$ \gcmcuad, with the total mass of pebbles that drifted after 3 Myr is 90 \ME \,. Also here shown are the \textit{enriched} (light-blue) and \textit{non-enriched} (red) cases and the crossover time is indicated for each case.} 
\label{pebbles_other_param}
\end{center}
\end{figure}

Figure \ref{pebbles_other_param} shows the sensitivity of the results to the assumed disk mass. If we double the disk mass, the formation timescales are shorter because the flux of pebbles is larger, producing larger accretion rates of pebbles. In this case, a total amount of 365 \ME \, of pebbles drift within 3 Myr and the likelihood of forming a mini-Neptune is much smaller than in the baseline LJ14 disk, yielding typically, mini-Neptunes' occurrence rates of \textit{zero}. 
On the other hand, for a disk with half of the mass of that in LJ14, the flux of pebbles diminishes, leading to a slower growth. In this case it is only possible to form mini-Neptunes by accounting for envelope enrichment, but the parameter space that leads to mini-Neptunes formation is reduced compared to the baseline disk model. For this case the total amount of pebbles that drifted after 3 Myr is 90 \ME, and it is only possible to form mini-Neptunes for very low-turbulent disks. 

We find that for $\Sigma_0 = 250$ \gcmcuad \, and $Z_0 = 0.005$ (same disk as in Fig.\ref{pebbles_other_param}b, but half of the metallcity), the pebbles mass that drifted after 3 Myr is only 30 \ME \,, and in this case, mini-Neptune formation does not occur, leaving lunar- to Earth-mass embryos. We find that in order to have a non-zero occurrence rate of mini-Neptunes at least $\sim$ 50 \ME \, of pebbbles must drift during the disk's lifetime.

\begin{figure}
\begin{center}
	\includegraphics[width=0.65\columnwidth]{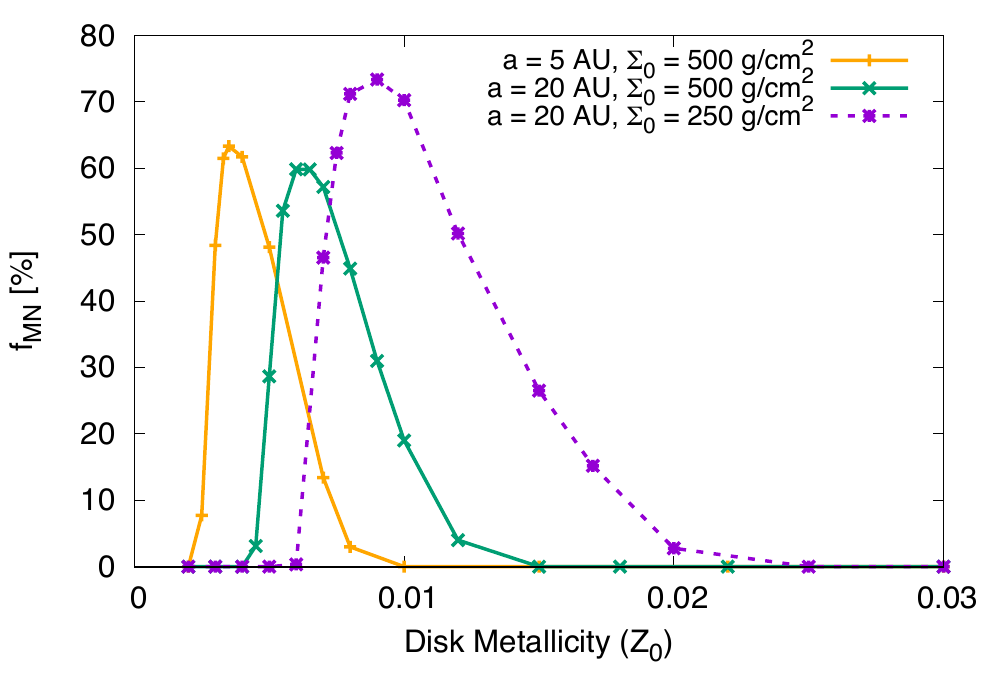}
	\caption{Mini-Neptune occurrence rate  (\textit{extended} definition) as a function of disk metallicity ($Z_0$) for the case of pebble accretion. $\alpha=10^{-5}$ and $p=1$. The orange ($a = 5$ AU) and green ($a = 20$ AU) curves correspond to $\Sigma_0 = 500$ \gcmcuad (baseline disk model of LJ14), whereas the violet corresponds to a disk with half of the mass ($\Sigma_0 = 250$ \gcmcuad) and $a = 20$ AU.} 
\label{fMN_Zdisk}
\end{center}
\end{figure}

If instead of changing the disk's mass we just modify the disk metallicity, we find that the peak in mini-Neptune occurrence rate takes place for disk metallicities \textit{lower than solar} for a typical disk as the baseline of LJ14, which contains an initial mass of solids of $\sim$ 100 \ME \, between 5 and 100 AU. The results are illustrated in Fig.\ref{fMN_Zdisk}, for $\alpha = 10^{-5}$, where accretion is 2D for most of the growth. The fast drop to a mini-Neptune occurrence rate of  0\% at $Z_0 \approx 0.004$ (for $a=20$ AU) is because the accretion becomes 3D. The reason why the formation of mini-Neptunes does not require high metallicity is that the optimal accretion rate of solids to obtain mini-Neptunes is of $\dot{M}_z  \sim 10^{-6}$ \ME/yr. This value is relatively low, so typical disks with metallicities equal or larger than solar are more prone to produce gas giants than mini-Neptunes. 

Figure \ref{fMN_Zdisk} also shows the mini-Neptune occurrence rate for a disk with half the mass of the baseline of LJ14 ($p = 1$, $\Sigma_0 = 250$ \gcmcuad). In this case the peak in mini-Neptune occurrence rate shifts towards a bit larger disk metallicity ($Z_0 \approx 0.009$) because when the total mass of the disk is reduced, the flux of pebbles diminishes, so a larger dust-to-gas ratio is required to have solid accretion rate of $\dot{M}_z  \sim 10^{-6}$ \ME/yr. Still, this disk metallicity is lower than solar, having an initial amount of solids of  $\sim$ 50 \ME \, between 5 and 100 AU.

In summary, there is "sweet spot" for mini-Neptune formation in the case of pebble accretion, which is a disk with a total amount of drifting pebbles in the range of 50 - 300 \ME, or more precisely, for $\dot{M}_z  \sim 10^{-6}$ \ME/yr. This condition is more likely to exist in low metallicity disks, and low disk viscosity.
If the disk is very turbulent (i.e., $\alpha > 10^{-3}$), the accretion rate of solids is too low, and gas accretion is unlikely to take place during the disk's lifetime, so in this case we are left with embryos of lunar- to Earth-mass.

\subsubsection{Planetesimals}\label{pla_general}
\begin{figure}
\begin{center}
\includegraphics[width=0.8\columnwidth]{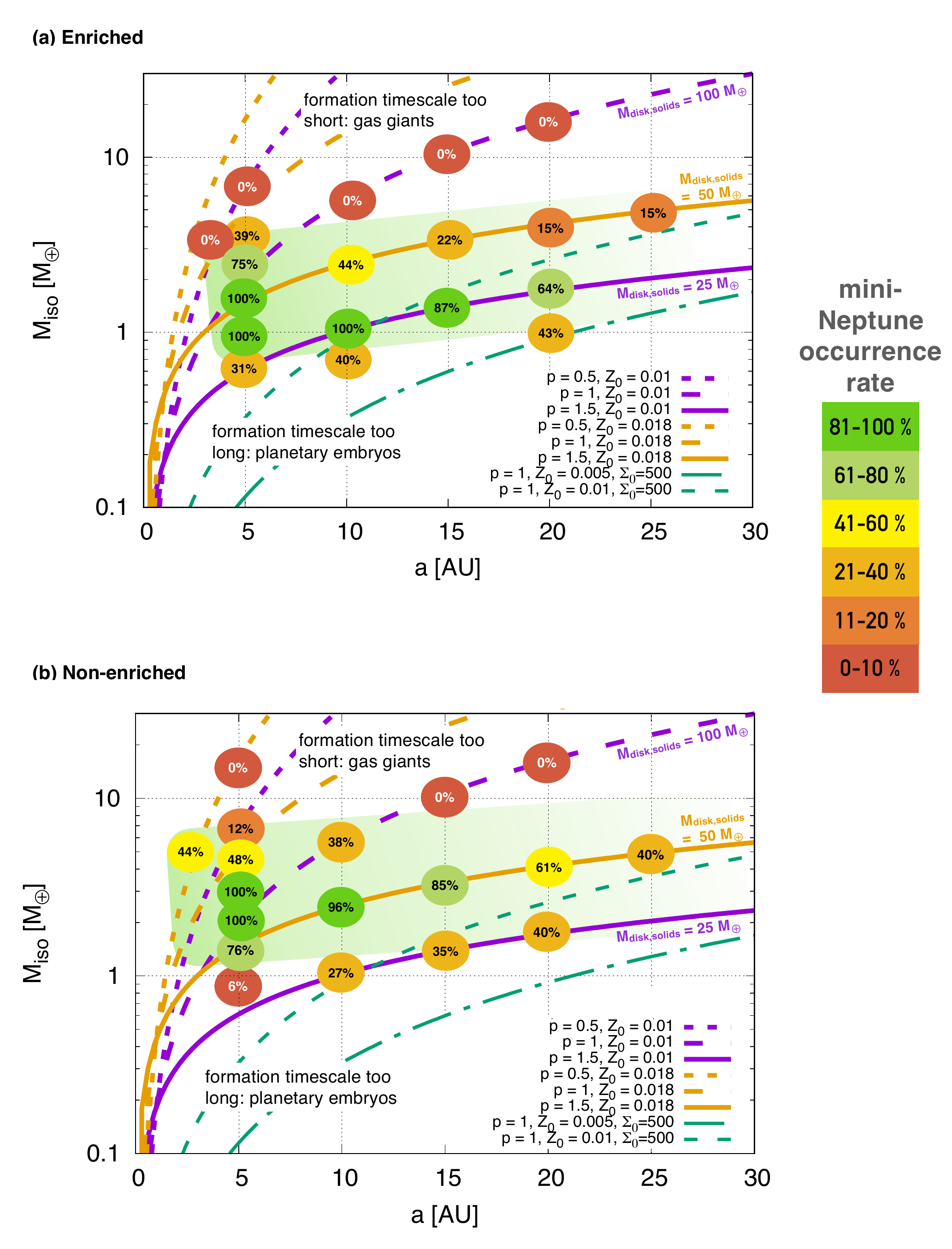} 
\caption{The derived occurrence rate of mini-Neptunes (\textit{extended} definition) for different semimajor axis ($a$) and isolation mass ($\Miso$) for the case of planetesimal accretion. The curves correspond to different disk parameters as indicated in the figure (if $\Sigma_0$ is not stated it is taken as 1700 \gcmcuad, as in the baseline model), although the occurrence rate is found to depend solely on $a$ and $\Miso$. Above each curve we indicate the total mass of solids between 5 and 30 AU for each disk model. The top and bottom panels correspond to the \textit{enriched} and \textit{non-enriched} cases, respectively. The green-shaded areas indicate the "sweet-spot" for the formation of mini-Neptunes.}
\label{maps}
\end{center}
\end{figure}

The results for planetesimal accretion for different disk models are shown in Fig.\ref{maps}. 
We corroborate that the only factors determining the occurrence rate of mini-Neptunes are $\Miso$ and $a$. In other words, if we change $\Sigma_0, Z_0$ or $p$ in order to have the same isolation mass at the same semimajor axis, 
we obtain exactly the same occurrence rate (of any type of planet).
Figure \ref{maps} shows that there is a "sweet spot" for the formation of mini-Neptunes, and that the mini-Neptune "occurrence region" is shifted to smaller $\Miso$ with envelope enrichment. 
This happens because gas accretion starts sooner when envelope enrichment is included. 
The general trends of the mini-Neptune occurrence rate can be understood as follows.

\begin{enumerate}
\item Fixed $a$: for small $\Miso$, formation timescales are so long that gas accretion is negligible during the disk's lifetime. There is a threshold for which $\Miso$ is large enough to allow gas accretion in the first 3 Myr.  For high values of $\Miso$, phase-2 becomes so short that the planet becomes a gas giant, leading for the occurrence rate of mini-Neptune to drop for large $\Miso$.

\item Fixed $\Miso$: the larger the semimajor axis is, the longer is the duration of phase 1 \citep{P96}. Thus, the sweet spot for mini-Neptunes (always during phase-2) occurs for times larger than 3 Myr as we increase $a$, which decreases the mini-Neptune occurrence rate.  
\end{enumerate}

\subsubsection{Lower accretion rates of planetesimals}
We find that if we implement lower planetesimal accretion rates than the one given by Eq.\ref{rafikov} (like the \textit{moderate accretion rates} described in \citet{Rafikov11}, for $\sim$10 km planetesimals), phase-1 lasts 3 Myr with the surface densities given by the baseline MMSN disk. During phase-1 gas accretion is negligible, therefore the occurrence rate of both mini-Neptunes and Neptunes is negligible. If we invoke larger surface densities of solids  (2-3 times the MMSN) phase-1 is reduced, but still, it lasts at least 1.6 Myr (at 5 AU). For this most favorable case we get occurrence rates if mini-Neptunes of 7\% for a 2-MMSN disk, and of Neptunes of 5\% for a 3-MMSN.

\subsection{The Formation of Neptunes}\label{neptunes}
We find that Neptune-like planets are not a common output both for the pebble and planetesimal cases. 
With low dust opacities and planetesimal accretion, for all the models with isolation masses above the green-shaded areas of Fig.{\ref{maps}}, either the duration of phase-2 is too short, or the amount of H-He when the planet crosses the $10 \leqslant \Mp \leqslant$ 20 \ME \, is too large. Hence, the most likely outcome is a gas giant, not a Neptune. 

Indeed, with low dust-opacities, as presented in Fig.{\ref{maps}}, the occurrence rate of Neptunes is \textit{zero}, with the exception being cases with embryos located at 25 AU, which reach Neptune occurrence rate of 10\%.  A necessary condition to get a non-zero occurrence rate of Neptunes is to get  10 <$\Miso$ < 15 \ME \, (consistent with a surface density of solids equivalent to 2-3 MMSN) and a relatively long phase-2. This is the reason why with low dust opacities formation is not possible, because for the mentioned isolation masses,  the longest derived duration of phase-2 is only 0.6 Myr. Assuming that the planet crosses the "Neptune region" during phase-2, this would yield a maximum occurrence rate of Neptunes of 20\% (at 5 AU).  

The only way to have Neptune occurrence rates relatively large with planetesimal accretion is without envelope enrichment and high-dust opacities, although the feasibility of such high dust opacities requires further investigations. When high dust opacities are invoked, a maximum occurrence rate of Neptunes of $\sim$ 40\% is found for $5 < a < 10$ AU, $p = 1$ and $Z_0 = 0.018$, or  $p = 0.5$ and $Z_0 = 0.01$; i.e, disks with masses of solids in the range of 200-400 \ME \, between 5 and 30 AU. Under similar conditions, \citet{Guilera11} are able to form Uranus and Neptune simultaneously. 

With pebble accretion there is no intermediate-mass long phase, so we find as well very low occurrence rate of Neptunes, typically zero, or $\sim$ 10\% in the best cases (see Tables \ref{tab1} and \ref{tab2}). We find that the combination of high dust opacities and envelope enrichment enhances slightly the possibility of the formation of Neptunes (see Table \ref{tab2}, pebble accretion with enrichment at $a = 5$ AU).

A possible mechanism to form Neptunes is the merging of mini-Neptunes, as proposed by \citet{Izidoro15b}. It was shown that by colliding 3-6 \ME \, embryos from a total mass of solids in embryos of 30-60 \ME, Uranus and Neptune can form. Our calculations show that mini-Neptunes are relatively easy to form under the same assumption of amount of solids. Therefore, we provide the seeds needed to form Neptunes. 
Thus, a key question that remains to be answered is what determines the formation of mini-Neptunes vs. Neptunes in planetary systems. 

It is interesting to note that for the cases where we manage to form Neptune-like planets, the embryos are located at radial distances $a<10$ AU, smaller than the locations of Uranus and Neptune in our Solar System. We therefore suggest that also in the pebble accretion case Uranus and Neptune cannot form in situ as argued by \citet{Lambrechts14}. Instead, our results are consistent with the scenario in which the two planets formed at smaller radial distances followed by outward migration \citep{NiceModel}.

\subsection{Formation inside the iceline}\label{inside_iceline}
Several studies explored the possibility of in-situ formation of super-Earths and mini-Neptunes inside the iceline, assuming planetesimal accretion \citep{IkomaHori12, Inamdar15,BodLiss14}. 
All of these investigations find that  in-situ formation is rather unlikely. The reason for this is illustrated in Fig.\ref{maps}: the isolation masses are extremely small. Indeed, we find that for planetesimal accretion in-situ formation at $a = 1 $ AU assuming 1 MMSN yields an isolation mass of $\Miso \approx 0.15 $ \ME. 
The envelope mass at this stage is hardly $\Menv = 2 \times 10^{-7} \, \Mearth$. Rocky planetesimals are expected to be fully disrupted before reaching the core when $\Menv \sim 0.01 \, \Mearth$ \citep{Mordasini06}. 
Therefore, unless some fast formation mechanism to start with an Earth-mass embryo is invoked \citep{BodLiss14}, it is not possible to form a few Earth-mass planet with a considerable amount of H-He in mass inside the iceline if the dominant accretion process is that of planetesimals.

Regarding growth with pebbles, the simple model of LJ14 is probably not valid inside the iceline. 
First, the dust properties might vary since the composition and the sizes of the grains change. 
Second, inside the iceline the solids tend to pile-up \citep{Drazkowska16}, so the assumption of a continuous flux of pebbles is not applicable. 
Concerning the first challenge, \citet{Morby15} showed that the growth of an embryo with pebbles inside the iceline is extremely inefficient as long as icy pebbles disrupt when crossing the iceline and release small dust grains. A Mars-size object is expected to grow from accretion of silicate dust grains (just like with planetesimal accretion), but not super-Earth mass planets. On the other hand, \citet{Drazkowska16} finds that rocky planets could form in-situ by pile-ups of solids (up to 1000 M$_{\oplus}$ at 1 AU). These objects could accrete some H-He and be enriched afterwards by silicate-grains, albeit gas accretion inside the iceline is challenging, given that hot gas tends to flow rather to stay bound to the planet \citep{Ormel15}.

\subsection{Dependence on disk's lifetime}\label{lifetime}

In all the results presented above, the disk's mean lifetime was set to be 3 Myr and the planetary occurrence rates were computed assuming the disk disappears after 1-10 Myr. In this section we explore the dependence of the emergence of mini-Neptunes on these parameters. 
We consider two cases: reducing the disk mean lifetime by half or increasing it by a factor of 2.  If we consider $\tdisk = 1.5$ Myr instead of 3 (and same $t_0$ and $t_M$), then for the baseline enriched case and $a=20$ AU with pebble accretion, the occurrence rate of mini-Neptunes rises from 41\% to 84\%. This happens because the flux of pebbles (which is proportional to exp(-time/$\tdisk)$) decays more slowly for shorter disks' lifetime. 
Therefore, the solid accretion rate is smaller and the protoplanet can spend more time in the mini-Neptune region before becoming a gas giant.  The opposite occurs when we set $\tdisk = 6$ Myr: the occurrence rate of mini-Neptunes falls now to 23\%. 

For the same baseline case for planetesimals, i.e., enriched at $a=20$ AU, the occurrence rates increase in the same direction as with pebbles: for $\tdisk = 1.5$ Myr, $f_{MN} = 27$ \% instead of 15\%; and $f_{MN} = 10$ \%  $\tdisk = 6$ Myr.  

It is interesting to note that for the cases where $\tdisk = 1.5$, if we narrow the range of possible disks' lifetimes to 1-3 Myr, then the occurrence rate of mini-Neptunes is 100\% for pebble accretion and 37\% for planetesimal accretion (for baseline model at $a=20$ AU, enriched cases). According to \citet{Ribas15}, systems with stellar masses larger than $\sim$ 2 \Msun show no presence of disks for ages larger than 3 Myr. In this context, the formation of mini-Neptunes would be favored for such systems. 

As we illustrate with the examples of this section, it is clear that  the likelihood to obtain a mini-Neptune depends on the disk's lifetime. Therefore, constraining disk lifetimes form observations will allow us to provide more accurate  predictions of mini-Neptune's occurrence rates from formation models. Viceversa, when combined with statistics of mini-Neptune occurrence rates, the models presented here could be used to constrain the origin of the observed systems.
Note, however, that due to the possibility of migration, tracking the origin of the planet is not trivial.

\subsection{Do mini-Neptunes survive migration?} 
Although in-situ formation of mini-Neptunes is rather unlikely (see Sect. \ref{inside_iceline}), most of the detected exoplanets orbit within about 1 AU from the central star. As a result, it is important to test whether mini-Neptunes are a common outcome also when migration is included. 
In order to do so, one needs to couple the formation models with consistent migration rates and the feedback on the disk, which is beyond the scope of this work. 
Nevertheless, by invoking simple timescale arguments we can estimate the effect of migration.  Type-I migration timescales are typically of $ \tau_{\text{mig}}\sim10^5$ years \citep[e.g.,][]{Tanaka02}. This is so short that planets that survive must have started to migrate late in the lifetime of the disk, when most of the solids had already either accreted to form larger objects or were lost into the star. 
Even if small planetesimals/pebbles exist, the planet is unlikely to accrete many solids during this short timescale. 
While this argument suggests that the planet is not expected to accrete many solids, it can still accrete gas. 
Nevertheless, our simulations show that for the favorable cases of mini-Neptune formation, when solid accretion stops, the timescale to accrete H-He is of the order of $10^6$ years.  This timescale is in fact a lower bound because there is no heating from the bombardment of solids that prevents the planet from a fast cooling and contraction. We therefore conclude that the mini-Neptunes formed beyond the iceline are unlikely to  accrete substantial H-He while migrating inwards, remaining as mini-Neptunes.

\section{Discussion and conclusions}

We investigate the feasibility of forming mini-Neptunes in different environments with an emphasis on the solid accretion rate, formation location, envelope metallicity and opacity. 
We find that atmosphere enrichment enhances the probability of forming mini-Neptunes. 
Despite the large uncertainties in the physical properties of the disk, and the planetary formation histories, we show that the formation of mini-Neptunes is efficient for various disk parameters.  
We find a "sweet spot" for mini-Neptune formation beyond the iceline for low-mass disks of solids ($\sim$30-200 \ME) - this can correspond either to low disk masses or to low metallicities. 

It should be noted that the mini-Neptunes detected so far have short periods, and therefore, according to our simulations, should have migrated from beyond the iceline. In addition, they could have lost part of their envelopes due to photoevaporation. As a result, the H-He masses we derived here should be taken as upper bounds. 
It is interesting to note that for the cases with envelope enrichment the cores are purely rocky (as the water remains in the envelope). A recent re-analysis of the Kepler data suggests that small planets have a maximum radius of 1.6 \RE \, \citep{Fulton17}, which corresponds to the maximum radius of a pure rocky object \citep{Rogers15}. This finding supports the scenario of atmospheric photoevaporation of low-mass planets leading to "naked rocky cores" assuming most of the oxygen is also lost. Although in this work we have not included this effect, it is consistent with our enrichment cases where the volatiles remain in the envelope and do not sink to the core.

For pebble accretion, the occurrence rate of mini-Neptunes without envelope enrichment is typically very low, while with envelope enrichment, which is the most likely scenario, the occurrence rate is between 10\% and 80\% (with pebble masses  between 50 and 200 \ME). In addition, we suggest that the output of planet formation is very sensitive to the disk turbulence. In very turbulent disks ($\alpha > 10^{-3}$), pebble accretion can be very inefficient, forming lunar- to Earth-embryos and not gas-rich planets. If several of these embryos are formed, consequent collisions can lead to the formation of larger terrestrial planets. We find that mini-Neptune formation is preferable for relatively low solid accretion rates ($\sim 10^{-6}$ \ME/yr) which typically corresponds to low metallicity environments.

For planetesimal accretion, the formation of mini-Neptunes is possible when the planetesimals are small ($\sim$100 m).  With envelope enrichment the isolation mass (total heavy-element mass) is smaller than in the non-enriched case.  When comparing pebble with planetesimal accretion, we find that the in the former the occurrence rate of mini-Neptunes is higher further out ($a\sim30$ AU), whereas with planetesimals, it is typically higher for smaller semimajor axes ($a\sim5$ AU) although the actual numbers also depend on the assumed opacity.

Finally, we find that Neptunes are very difficult to form, with pebbles or planetesimals, when low opacity models are implemented. The only cases with relatively high occurrence rate of Neptunes ($\sim 40 \%$) are for planetesimal accretion, and disks with larger amounts of solids (i.e., 2-3 MMSN disk) coupled with larger dust opacities. This may suggest that Neptunes in low metallicity environments form by  merging of mini-Neptunes \citep{Izidoro15b}.

\bigskip

\textbf{Acknowledgements.}
We thank the anonymous referee for valuable comments.
This work has been carried out within the framework of the National Centre for Competence in Research PlanetS, supported by the Swiss National Foundation. 
R.H. acknowledges support from SNSF grant 200021\_169054.

\newpage

\begin{deluxetable*}{ccccccccc} 
\tablenum{1}
\tablecaption{Summary of results comparing planetesimal and pebble accretion for low dust opacities. \label{tab1}}
\tablewidth{0pt}
\tablehead{
\colhead{run} & \colhead{a [AU]} & \colhead{Enrichment} &
\colhead{$t_{\text{cross}}$ [Myr]} & \colhead{$t_i$ [Myr]}  & \colhead{$t_f$ [Myr]} & \multicolumn2c{f$_{MN}$} & \colhead{f$_{\text{Nept}}$}  \\
 & & & & &  &  restricted  & extended &
}

\startdata
pebbles & 5 & No & 1.9 & 1.24 & 1.27 & 0\% & 1\% & 15\%\\
 &  & Yes & 0.65 & <1 & <1 &  0\% & 0\% & 0\%  \\
planetesimals & 5  & No & 53 & 1.74 & >10 &  33\% &  76\% & 0\%  \\
                      &   & Yes & 14 & <1 & >10  & 83\% &  100\% & 0\%  \\
pebbles & 20 & No &  4.4 & 4.5 & 7.2 &  9\% & 19\% & 0\% \\
 	     &    & Yes & 1.8 & <1 & 2.5 & 27\% & 41\% & 0\% \\
planetesimals & 20 & No & 6 &  1.7 &  5.6  & 23\% &  61\% & 0\%\\
                       &   & Yes & 1.5 & <1 &  1.47  & 6\% & 15\% & 0\% \\
\enddata
\tablecomments{Results for the MMSN disk with dust opacities of \citet{Mordasini14} with $\tcross$ being the crossover time, $t_i$ and $t_f$ the time at which the planet enters and exits the \textit{extended} mini-Neptune regime, respectively;  f$_{MN}$ the mini-Neptune occurrence rate (see Sect.\ref{method}) and $f_{\text{Nept}}$ the occurrence rate of Neptunes (computed using as well Eq.\ref{eq_occ_rate} but considering the times of ingress and egress of the Neptune region). For the cases of pebble accretion the viscosity parameter is set to $\alpha = 10^{-5}$. For the case of planetesimals, the initial surface densities of solids at 5 and 20 AU are 2.7 \gcmcuad  and 0.34 \gcmcuad, respectively.}
\end{deluxetable*}

\begin{deluxetable*}{ccccccccc} 
\tablenum{2}
\tablecaption{Summary of results comparing planetesimal and pebble accretion for high dust opacities. \label{tab2}}
\tablewidth{0pt}

\tablehead{
\colhead{run} & \colhead{a [AU]} & \colhead{Enrichment} &
\colhead{$t_{\text{cross}}$ [Myr]} & \colhead{$t_i$ [Myr]}  & \colhead{$t_f$ [Myr]} & \multicolumn2c{f$_{MN}$} & \colhead{f$_{\text{Nept}}$}  \\
 & & & & &  &  restricted  & extended &
}

\startdata
pebbles & 5 & No & 2.7 & - & - &  0 \% & 0\% & 0\% \\
 	     &    & Yes & 1.5 & <1 &  1.04   & 1\% & 6\% & 5\% \\
planetesimals & 5 & No & 650 & - & -  & 0 \% & 0 \% & 0\%   \\
                      &  & Yes & > 4 & - & -  & 0 \% & 0 \%  & 0\% \\
pebbles & 20 & No & 27 & 11 & 27 &  0\% & 0\% & 0\% \\
 	     &      & Yes & 6.3 & 1.3 & 6.6  & 40\%  & 79\% & 0\% \\
planetesimals & 20 & No & 72 &  6.17 & >10  & 0 \% & 14\% & 0\%  \\
                       &    & Yes & 23 & 1.57 & >10  & 66 \% & 82\% & 0\%  \\
\enddata
\tablecomments{Same as Table \ref{tab1} but for high dust opacities (i.e., 100 $\times$ \citet{Mordasini14}).}
\end{deluxetable*}

\end{document}